# High Noise Immune Time-domain Inversion via Cascade Network (TICaN) for Complex Scatterers

Hongyu Gao, Yinpeng Wang, Qiang Ren, Zixi Wang, Liangcheng Deng and Chenyu Shi

*Abstract*—In this paper, a high noise immune time-domain inversion cascade network (TICaN) is proposed to reconstruct scatterers from the measured electromagnetic fields. The TICaN is comprised of a denoising block aiming at improving the signal-to-noise ratio, and an inversion block to reconstruct the electromagnetic properties from the raw time-domain measurements. The scatterers investigated in this study include complicated geometry shapes and high contrast, which cover the stratum layer, lossy medium and hyperfine structure, etc. After being well trained, the performance of the TICaN is evaluated from the perspective of accuracy, noise-immunity, computational acceleration, and generalizability. It can be proven that the proposed framework can realize high-precision inversion under high-intensity noise environments. Compared with traditional reconstruction methods, TICaN avoids the tedious iterative calculation by utilizing the parallel computing ability of GPU and thus significantly reduce the computing time. Besides, the proposed TICaN has certain generalization ability in reconstructing the unknown scatterers such as the famous "Austria" rings. Herein, it is confident that the proposed TICaN will serve as a new path for real-time quantitative microwave imaging for various practical scenarios.

## I. INTRODUCTION

ELECTROMAGNETIC inverse problems (ISPs) aim to reconstruct the nature of unknown scatterers, including their shapes, positions and electromagnetic properties, through the scattered fields collected by external receivers. Nowadays, these techniques have been widely applied in remote sensing [1], prospecting [2], biomedical imaging [3][4] as an accurate and nondestructive detection tool [5][6].

Generally, the nonlinearity and ill condition are two intrinsic challenges widely encountered in the ISPs. Among them, the nonlinear operators require more computing resources, thus enhance the calculating complexity, while the ill condition usually attributes to underdetermined equations which result in multiplicity. Consequently, it is tough to acquire a robust and precise solution in the ISPs for years. For weak scattering scenarios, linear approximation is often utilized, which has the merit of fast computing and low resource consumption. Classic linear inversion algorithms comprise Born approximation (BA) [7] and Kirchhoff approximation (KA) [8], which is adopted for dielectrics and conductors respectively. However, when the scattered fields turn stronger, iterations are indispensable during the inversion process. Traditional frequency domain iterative algorithms include Born iteration method (BIM) [9], contrast source inversion method (CSI) [10], the subspace optimization method (SOM) [11], and etc. Compared to frequency approaches, time-domain imaging algorithms operate directly on time domain data. Exiting iteration approaches in time domain contain forward backward time-stepping (FBTS) [12], time-domain distorted-Born iterative method (TD-DBIM) [13], and time-domain Gauss-Newton inversion (GNI) [14]. In these methods, a finite difference time domain (FDTD) algorithm is usually applied as the forward solver [15]. In order to retrieve the image of the scatterer, the FDTD based forward modeling tool is employed in each iteration. Consequently, one of the major drawbacks in these iterative methods is that they are time-consuming hence not suitable for real-time inversion. Besides, these methods need a large quantity of computing source to complete the high complexity and high-nonlinear iterative calculation.

In recent years, the Deep Neural network (DNN) has achieved extensive advance in computational physics. Given abundant labeled data and sufficient computing units, the DNN is capable of revealing the inherent characteristics of the target system through the training process. For forward problems, several research works via DNN have been investigated to solve physical fields related to electromagnetics [16]-[19], optics [20], fluids [21] and heat transfer [22]. Very recently, approaches based on deep learning with regression features have also been employed on ISP successfully. For instance, Wei *et al.* [23] proposed three inversion schemes based on U-net CNN to solve nonlinear ISPs and demonstrated that the dominant current schemes (DCS) outperform the other two in computational complexity and accuracy. Li *et al.* [24] develop a DL framework cascaded of multilayer complex-valued residual convolutional neural network modules (DeepNIS) to establish an end-to-end map from a rough image from the back-propagation (BP) method to a fine permittivity image, greatly reducing the calculation time compared to traditional

This work was supported in part by the National Natural Science Foundation of China (Grant No. 92166107), and in part by Defense Industrial Technology Development Program through (Grant No. JCKY2020601B011), and in part by Stable Operation Project, and in part by the National Science and Technology Major Project (Grant No. 2019-VIII-0009-0170). (Corresponding authors: Qiang Ren.)

The authors are with the School of Electronic and Information Engineering, Beihang University, Beijing, 100191, China (e-mail: qiangren@buaa.edu.cn).

algorithms. Xu *et al.* [25] brought forward an inversion approach to solve high-nonlinear ISPs with only phaseless data by the U-net convolutional neural network and demonstrate that the proposed two training schemes perform better from the perspective of accuracy, generalization ability and robustness. Yao *et al.* [26] proposed a two-step DL based approach consisting of deep convolutional networks to retrieve initial permittivity from measured scattering data and then refine the reconstruction images. Additionally, Lin *et al.* [27] presented a DL based low-frequency scheme to resolve inverse scattering problems and alleviated the nonlinearity by drawing into the LF component measurements. Moreover, Ye *et al.* [28] proposed a GAN network as an encoder-decoder structure configured with the attention scheme to reconstruct objects embedded in inhomogeneous background with the input rough image generated by the distorted-Born backpropagation. Furthermore, Xiao *et al.* [29] adopted a dual-module scheme, consisting of an extreme learning machine (ELM) and a nonlinear mapping module (NMM), to successfully reconstruct inhomogeneous scatterers with high contrasts and large electrical dimensions.

These emerging deep learning techniques overcome the deflects of conventional algorithms in low computing speed and high resource demanding, enormously promoting the inversion efficiency. However, the state-of-the-art deep learning frameworks mainly concentrate on the frequency-domain inversion, whose data acquisition is not as direct as the time domain. In addition, due to the lack of priori knowledge, multiple transmitting antennas with different frequencies and locations are often used in turn to produce sufficient information in these methods, which makes the inversion process more cumbersome. To this end, an elaborate designed cascade deep learning framework is introduced in this research to reconstruct the electromagnetic constitutive parameters directly from the time-domain measurements. Compared to the previous works, the proposed TICaN has the following innovations and advantages:

1) NOISE RESISTANCE: TICaN shows superiority when reconstructing the EM parameters under different types of noise with high intensity in real-site scenarios. In this research, Gaussian/Rayleigh/Uniform noises with the SNR as low as 1 dB are added to the received signals respectively to demonstrate the striking denoising ability.

2) TIME-DOMAIN DATA: TICaN employs directly the time-domain data which avoids the sophisticated Fourier transform in data preprocessing. Besides, since the wideband time-domain pulses apparently contains more information due to its wideband transient waveform, one transmitting antenna is sufficient to supply the corresponding inversion data, which undoubtfully simplify the data collection procedure.

3) COMPLEX SCATTERERS: The scatterers investigated in this research have intricate geometry shapes and high contrast (up to 9). Besides, more practical situations such as geological structure data set and lossy medium data set are analyzed in this paper. Due to the strong nonlinear effect, these datasets are unfriendly to the traditional full wave inversion algorithms.

4) HIGH PRECISION: TICaN can realize high-precision inversion with the resolution exceeding the Rayleigh limit. For hyperfine structures, the proposed network can obtain clearer reconstruction results than traditional iterative algorithms or existing networks.

5) REAL-TIME PREDICTION: A fully trained neural network model can give full play to the parallel computing ability of GPU and hence avoid repeated iterations. Consequently, it is possible so that little computing resource is required to reconstruct the dielectric image in real time.

This paper is organized as follows. In section II, the workflow of TICaN is briefly introduced firstly, followed by data generation process and the TICaN structure. In section III, test set is utilized to evaluate the performance of the proposed DL architecture from the perspectives of accuracy, noise-immunity, computational acceleration, and generalizability. In Section IV, conclusions are finally provided.

## II. METHODOLOGY

Detailed procedures of data generation and training are presented in Fig. 1. In order to acquire the time-domain waveform, randomly generated scatterers are sent to a self-developed FDTD forward solver which is erenow validated by the commercial software WCT [30]. The EM fields computed by the forward solver is then added with noise and fed to TICaN for training.

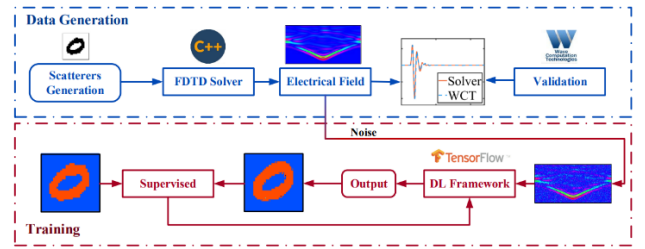

**Fig. 1.** The flowchart for TICaN to solve two-dimensional electromagnetic time-domain inversion problem. The forward solver based on FDTD is validated by the commercial software WCT. Various noises with different SNRs are added to the received waveform to form the data sets. The generated data along with the ground truth is fed to the TICaN for supervised learning.

### A. Scatterer

In order to yield a comprehensive assessment of the TICaN, six datasets with totally different geometry shapes and electromagnetic properties are included in the experiments. As displayed in Fig. 2, dataset B contains barely circles and rings, which merely verifies the fundamental inversion ability. Dataset C includes several rectangles, which is more challengeable since the straight boundary is more prone to blur during the reconstruction process. Dataset D is comprised of the handwriting numbers and letters (EMNIST [31]) which evaluates the inversion capability of the network to hyperfine structures. Datasets E is composed of multiple scatterers which properly assess the reconstruction performance of the network for complex geometric objects. Additionally, the dataset A based on the stratum layer models is also introduced to verify the application of the framework in the field of geoscience. The above scatterers are all lossless media. However, in some specific scenarios such as biomedical imaging, the loss caused by electrical conductivity in human tissue cannot be ignored. Therefore, it is indispensable to reconstruct the permittivity and conductivity simultaneously. To this end, we have introduced dataset F to validate the inversion ability of the network for lossy media.

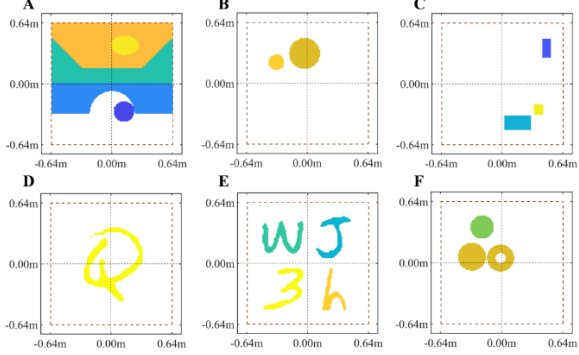

**Fig. 2.** Six datasets for TICaN. A: the stratum layer models, B: circles and rings, C: rectangles, D: handwriting numbers and letters from EMNIST, E: multiple scatterers from dataset D, F: lossy scatterers.

## B. Forward Solver

After generating the scatterers, an efficient and accurate forward solver is essential in constructing a large amount of data. In this research, the finite-difference time-domain (FDTD) [32] algorithm is applied with C language [33] to build the dataset. The FDTD method approximates the Faraday's law and Ampere's law in Maxwell's equations with a second-order central difference, which can be written as follows:

$$-\mu \frac{\partial \mathbf{H}}{\partial t} = \nabla \times \mathbf{E} \tag{1}$$

$$\mathbf{J}_s + \sigma \mathbf{E} + \varepsilon \frac{\partial \mathbf{E}}{\partial t} = \nabla \times \mathbf{H} \tag{2}$$

where $\mathbf{E}$ and $\mathbf{H}$ represent the electric and magnetic fields, while $\mathbf{J}_s$, $\sigma$, $\mu$ and $\varepsilon$ denotes the impressed currency, the conductivity, permeability and permittivity, respectively. For the convenience of calculation, we only consider a 2-D TM$_z$ mode. Thus, within a Cartesian coordinate system, scalar form of Maxwell's equations can be simplified as

$$-\mu \frac{\partial H_x}{\partial t} = \frac{\partial E_z}{\partial y} \tag{3}$$

$$\mu \frac{\partial H_y}{\partial t} = \frac{\partial E_z}{\partial x} \tag{4}$$

$$J_{sz} + \sigma E_z + \varepsilon \frac{\partial E_z}{\partial t} = \frac{\partial H_z}{\partial x} - \frac{\partial H_x}{\partial y} \tag{5}$$

By staggering the fields in space and time, the time domain forward simulation of the electromagnetic fields can be implemented. The configuration of computational region is presented in Fig. 3, where the scatterer is located in the center of the scattering area, and several observers are properly arranged along a circle. The size of the scatterers is within 75×75cm and 128 observers are placed 75cm away from the center. It is worth noting that there is only one transmitter, hence we only needs to receive the time-domain data once, which undoubtedly simplifies the data collection process. Here, each receiver obtains a signal with a length of 12.8 ns with 128 time steps, hence an electric field matrix of 128×128 (A square matrix will facilitate the performance of image processing operations of the network) is fed into the inversion network. By altering the scatterer, a large quantity of training data can be acquired with high efficiency.

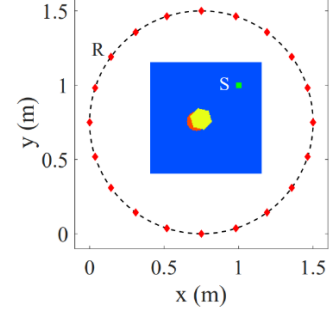

**Fig. 3.** Configuration of the computational region. The unknown scatterer is located at the center while the source and receivers are denoted as S and R, respectively.

Admittedly, for any electromagnetic simulation algorithm, numerical errors exist objectively. To ensure the credibility of the data set, the results of self-developed forward solver are compared to that of the commercial software WCT. The waveform at certain observer is presented in Fig. 4. As a result, the error of the electromagnetic field generated by the solver is extremely small, hence its accuracy can be verified.

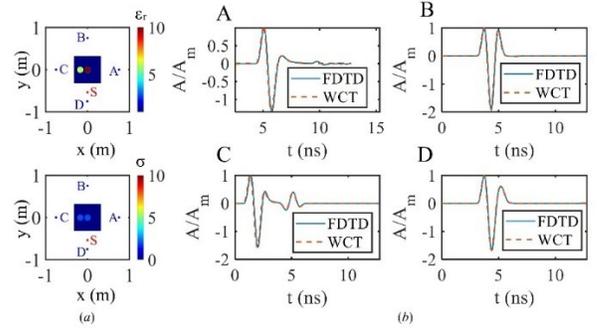

**Fig. 4.** Comparison between the self-developed forward solver and the commercial software WCT. (a) Configuration of the computational environment in the FDTD forward solver. The source is located at S, while A to F are six detectors used to receive the scattered field, and (b) electric field amplitudes calculated by the self-developed FDTD solver and WCT.

## C. Time-domain Inversion

The time-domain electromagnetic inversion can be regarded as an optimization problem whose objective functional is

$$F(p) = \frac{1}{2} \sum_{m=1}^{M} \sum_{n=1}^{N} \int_0^T \left\| E_{m,n}(p) - E_{m,n}^{mea} \right\| dt \tag{6}$$

where the inversion parameter to be reconstructed is the permittivity distribution. M and N indicate the numbers of the transmitters and the receivers while T represents the length of observation window. $E_{m,n}(p)$ and $E_{m,n}^{mea}$ stand for the calculated value (when the inversion parameter is supposed to be $p$) and the measured value. It should be noted that the subscripts $n$ and $m$ demonstrate the $n$-th receiver and the $m$-th transmitter. By introducing the Lagrange vector multipliers $\mathbf{e_m}$ and $\mathbf{h_m}$ corresponding to $\mathbf{E_m}$ and $\mathbf{H_m}$, we can transform the constrained minimization problem into an unconstrained minimization problem, whose augmented objective functional $F^a$ is

$$F^a(p) = \sum_{m=1}^{M}\int_0^T\int_V\left[\mathbf{e_m}\cdot\left(\nabla\times\mathbf{H_m}-\varepsilon_0\varepsilon_r\frac{\partial\mathbf{E_m}}{\partial t}\right)\right.$$
$$\left. +\mathbf{h_m}\cdot\left(\nabla\times\mathbf{E_m}+\mu_0\frac{\partial\mathbf{H_m}}{\partial t}\right)\right]\mathrm{d}V\mathrm{d}t+F(p) \tag{7}$$

Through variational, it is feasible to obtain the gradient of $F^a$ with respect to each component of $p$ in the reconstruction area:

$$g_\varepsilon=\frac{\delta F^a}{\delta \varepsilon_r}=-\varepsilon_0\sum_{m=0}^{M}\int_0^T\frac{\partial\mathbf{E_m}}{\partial t}\cdot\mathbf{e_m}dt \tag{8}$$

$$g_\sigma=\frac{\delta F^a}{\delta \sigma}=-\sum_{m=0}^{M}\int_0^T\mathbf{E_m}\cdot\mathbf{e_m}dt \tag{9}$$

where $\delta$ is the first order variational operator. Here, $\mathbf{e_m}$ and $\mathbf{h_m}$ satisfy:

$$\mu_0\frac{\partial\mathbf{h_m}}{\partial t}=\nabla\times\mathbf{e_m} \tag{10}$$

$$\sum_{n=1}^{N}\left[\mathbf{E_{m,n}}(p)-\mathbf{E_{m,n}^{mea}}\right]+\varepsilon_0\varepsilon_r\frac{\partial\mathbf{e_m}}{\partial t}-\sigma\mathbf{e_m}=-\nabla\times\mathbf{h_m} \tag{11}$$

It can be found that Eq. 11 and Eq. 2 has similar forms. To solve Eq. 11, repeated iterations of classical time-domain forward algorithms are indispensable which is impractical in real-time scenes. In order to avert cumbersome iterations, a deep learning framework TICaN is proposed, whose core idea is to use a group of neurons instead of the functional $F$ to obtain the optimal value of parameter $p$. In this paper, a DL framework is built for time-domain inversion, which has a quasi-U-net structure with Effecientnet $b_0$ as backbone. To better resist noise, a U-net for denoising is adopted before Effecientunet. The detailed introduction of each sub-module is presented as follows

*1) U-net for denoising*: In the ISP model, all the data utilized in the inversion process is obtained from the receivers outside the scattering region. Considering that the time-domain electric field acquired by detectors is easily affected by the external environment, it is valuable to recover a clean waveform $E_0(t)$ from a noisy observed $E(t)$. Three kinds of noise (Gaussian noise, Rayleigh noise and uniform noise.) are introduced to test the universality of denoising.

To achieve this goal, the U-net, widely used in image denoising [34][35], is employed to purify the received signal. The detailed architecture of the denoising framework is demonstrated in Fig. 5. This U-net utilized in denoising can be divided into two paths, the contracting path and the expansive path. The contracting path represents a series of downsampling processes, during which the number of channels doubles. It contains a total of 3 encoder blocks ended by an average-pooling layer. The expansive path is similar to contracting path, but the average-pooling layer is replaced by the upsampling layer, where the number of channels is halved, while the size of the feature map is doubled.

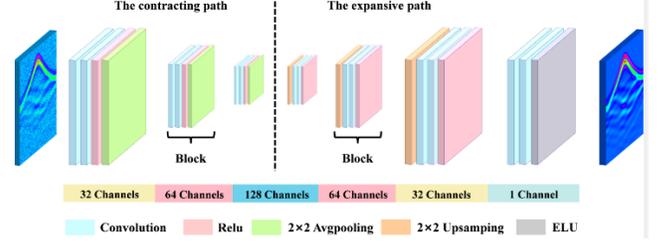

**Fig. 5.** The architecture and input-output scheme of the U-net in the denoising process. The whole structure can be divided into the contracting path and the expansive path, which represent a series of downsampling processes and upsampling processes respectively. The input is an image with the size of 256×256 representing the received waveform data, while the output is an image with the same size serving as the input of the Effecientunet for time-domain inversion.

The U-net is substantially suitable for denoising of two-dimensional images due to the following reasons:

1. During pooling and downsampling, double convolution layers allow the network to extract high-level features from low-level features and to expand the receptive field. In view of the weighted smoothing characteristics of the convolution kernel, a larger receptive field can effectively improve the filtering ability, thus improving the denoising performance.

2. Image filtering methods utilized in this research are mainly based on neighborhood filtering. Traditional filtering methods such as median filtering and mean filtering are all criticized by the loss of image details. However, the convolution layers in neural network can effectively preserve the feature map via neighborhood filtering, and U-net, as a fully convolutional network, is well suited for the denoising of two-dimensional images.

*2) Effecientunet for time-domain inversion*: Considering that the time domain waveform image of the scattered field can reflect the spatial distribution of the permittivity and conductivity to a certain extent, the full convolution network is selected for the reconstruction of the scatterers. This structure adopts the effecientnet-$b_0$ [36] as the backbone to replace the encoder of the U-net, which consists of several sub-blocks, including the stem block, head block and seven other blocks composed of several MobileConvoluotion (MBConv). In each MBConv, the Relu activation function caches the Swish activation function, and the SE module is used in the short connection part, which is similar to the residual link. The purpose of the SE module is to improve quality of network generation by explicitly modeling the mutual dependence between channels of convolution features, so that the network can learn to use global information to selectively emphasize informative features and suppress useless features.

In addition, DropConnect is used to replace the traditional drop method. The difference between DropConnect and Dropout is that in the process of training the neural network model, it is not a random discard of the output of the hidden layer node, but the input weight. This structure introduces the sparsity of the weight rather than the output vector of the layer, which enhances the generalization ability and averts overfitting.

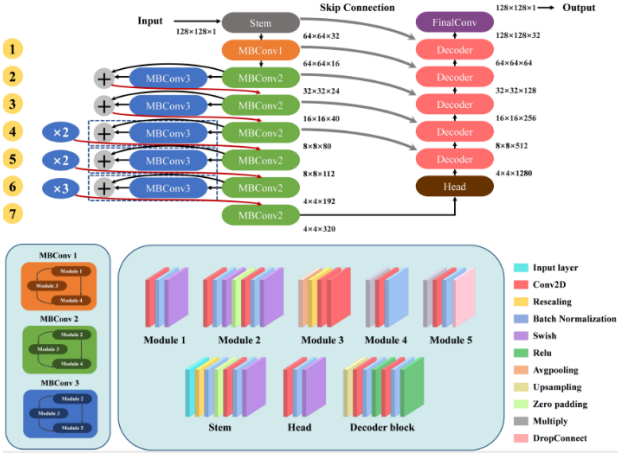

**Fig. 6** The structure of the Effecientunet. The structure consists of the stem block, head block and seven other blocks composed of several MobileConvoluotion (MBConv).

As shown in Fig. 6, each MBConv is composed of a series of modules. Module 1 is the starting point of seven sub-blocks while Module 2 is used for the other blocks. Module 3 is an SE module for short connection to Module 4 and Module 5, which connect the current MBConv to the other MBConv in a jump connection.

After feature extraction, Effecientnet-b0 realizes the reconstruction of permittivity and conductivity through a series of upsampling process. In order to prevent the gradient disappearance, skip connection is built between Efficientnet-b0 and the decoder. The starting point is the input layer while the batch normalization is the output of the first MBConv of the first four blocks, and the ending point is the output of the upsampling layer of the decoder, which is spliced in the channel direction.

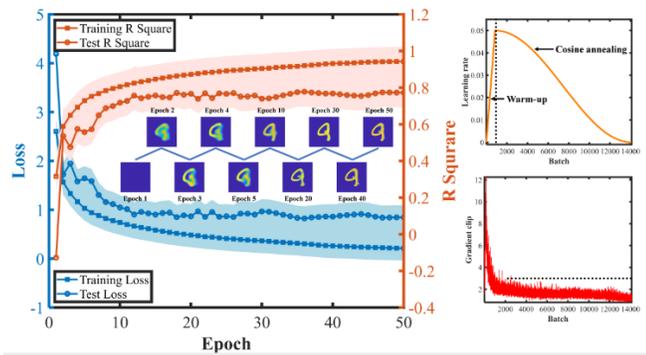

**Fig. 7** The loss and accuracy of the training process. (a) The training curve and the prediction results of each iteration step on the test set. (b) Specific techniques utilized in the training process.

Fig. 7 shows the loss and accuracy of the training process. Considering time-domain inversion is a continuous-valued regression prediction, the accuracy is measured by R Square. In addition, during the training process, Warm-up and Cosine annealing algorithms are used to adjust the learning rate, and gradient clips are inserted to determine whether the gradient explodes while $L_2$ regularization is used to improve the model performance. At last, the prediction results of each iteration step on the test set are shown in Fig. 7, which is similar to that of the traditional algorithm.

## III. RESULTS

In this section, the inversion results of the TICaN are analyzed detailly. Firstly, randomly selected scatterer field configurations from the six different test datasets are fed into the trained framework to reconstruct the respective scatterers. Next, the performance of the proposed DL architecture is evaluated from the perspectives of accuracy, noise-immunity, computational acceleration, and generalizability.

### A. Accuracy

At the beginning, several stochastically chosen reconstruction results are presented to intuitively demonstrate the prediction accuracy of the TICaN. Fig. 8 (i) and (iv) exhibit the ground truth and the reconstruction results of the scatterer while (v) is the absolute error of the two. It can be found that the proposed model inverts the shape, spatial distribution and permittivity of the scatterers with considerable fidelity and the error is practically negligible. In order to further illustrate the advantages of TICaN, we have drawn into the classical inversion algorithms DBIM [37] and SOM [38] as comparison whose results are presented in (ii) and (iii). Due to the high contrast of the target scatterer, the strong nonlinear effect is very significant greatly affects the inversion results. Although the traditional can generally reveal the geometric shape of scatterers, it cannot retrieve clear boundaries and detailed features. Based on the above results, it can be concluded that the proposed TICaN can retain considerable inversion imaging quality even facing sophisticated scatterers, which excel the traditional algorithm to a large extent.

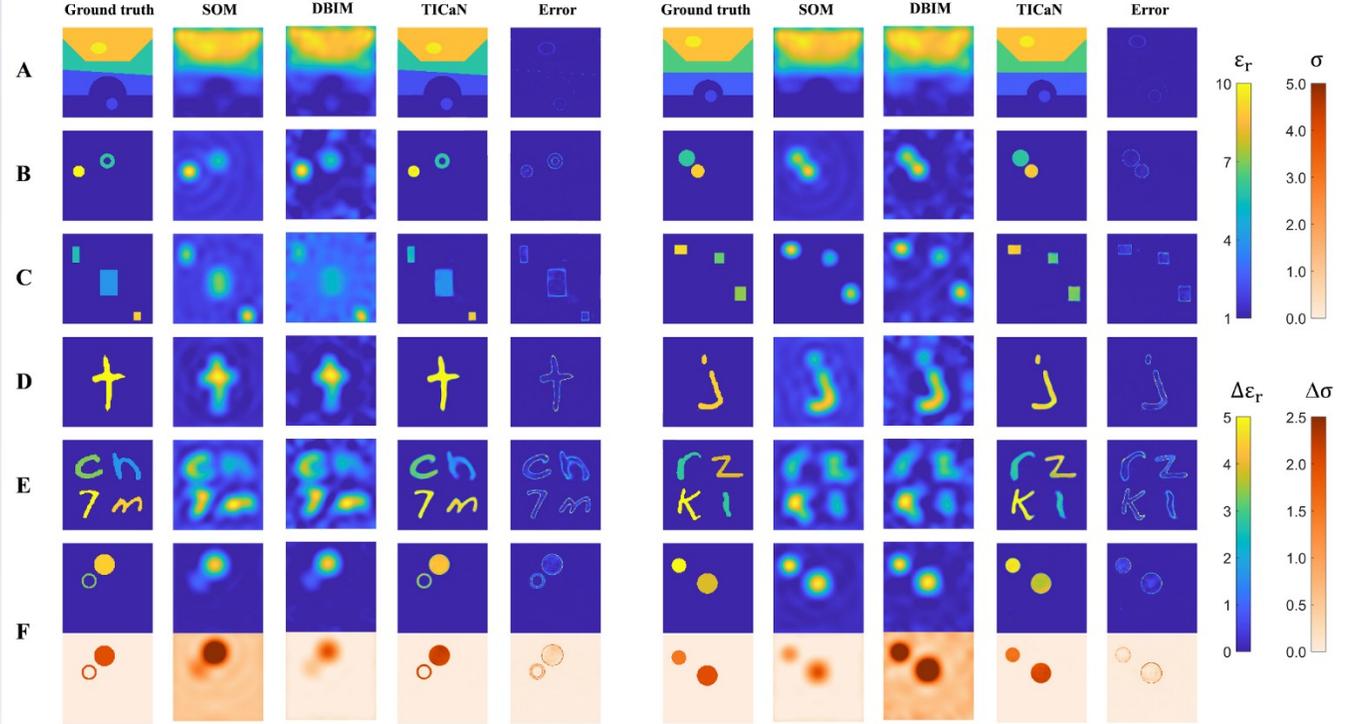

**Fig. 8** The reconstruction results of dataset A to F. (i) The ground truth, (ii) reconstruction results by SOM, (ii) reconstruction results by DBIM, (iv) reconstruction results by TICaN and (v) the error between (i) and (iv).

The above experimental results only qualitatively analyze the prediction results of the network. In order to further quantitatively measure the inversion accuracy, we introduce the PMSE, which is defined as:

$$PMSE = \frac{\sum_{j=1}^{J}\sum_{k=1}^{K}\left|g(j,k)-g'(i,k)\right|^2}{JKA^2} \quad (12)$$

where $g(j,k)$ and $g'(j,k)$ are the ground truth and predicted results by TICaN while A is the maximum value of the permittivity in the corresponding dataset. We have calculated the PMSE of the six datasets respectively and displayed it as a histogram in Fig. 9. It can be seen that most of the predicted results have a PMSE less than $10^{-1}$, which thoroughly emerges the robustness of the network.

### B. Noise-immunity

An impressive merit of the proposed TICaN is that it can be employed in high-noise environment. Therefore, it is necessary to validate the denoising ability of the proposed framework for diverse noise with different intensity. Here, the frequently encountered Gaussian noise, Rayleigh noise and uniform noise are investigated, whose results are displayed in Fig. 10. It can be found that the denoised wave agrees well with the ground truth, demonstrating the outstanding denoising performance of the U-net. To better quantify the denoising performance, the Peak Signal to Noise Ratio (PSNR) and Structural Similarity (SSIM) of 5000 samples with different SNRs are recorded, including data w/o denoising process. Since PSNR and SSIM are image quality indicators, the original waveform data is mapped to the pixel values from 0 to 255.

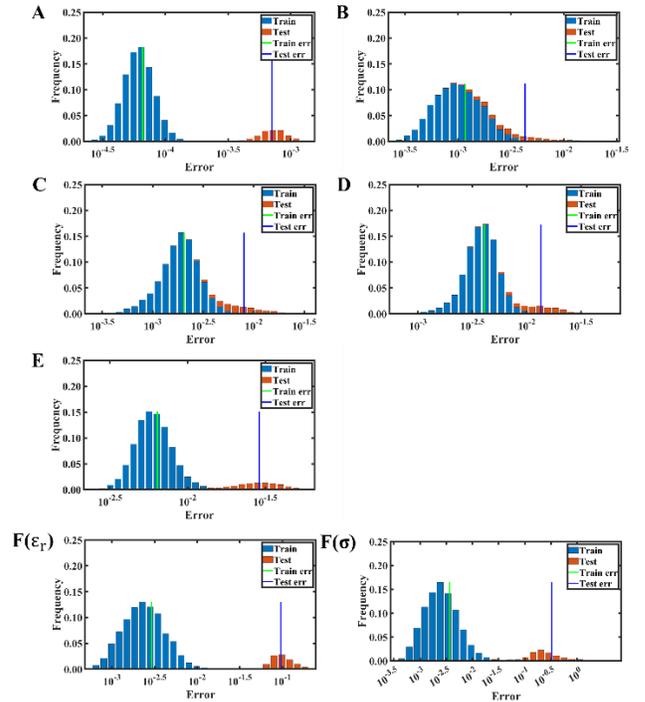

**Fig. 9** The PMSE of the six datasets on the training and testing set. A: the stratum layer models, B: circles and rings, C: rectangles, D: handwriting numbers and letters from EMNIST, E: multiple scatterers from dataset D, F: lossy scatterers.

| | 1 | | 3 | | 5 | |
|---|---|---|---|---|---|---|
| | **Before** | **After** | **Before** | **After** | **Before** | **After** |
| **Gaussian** | 27.58/0.47 | 46.43/0.99 | 29.58/0.57 | 47.27/0.99 | 31.57/0.66 | 47.31/0.99 |
| **Rayleigh** | 28.83/0.48 | 47.50/0.99 | 30.01/0.60 | 48.20/0.99 | 33.08/0.71 | 49.04/0.99 |
| **Uniform** | 28.30/0.48 | 46.88/0.99 | 29.84/0.60 | 48.81/0.99 | 33.01/0.71 | 48.98/0.99 |

Table 1 quantitatively presents the denoising capability of the network for noises with different types and SNR. It can be acquired that even for images with SNR of 1dB, U-net can reduce the noise to an ultra-low level with a PSNR of around 47dB and an SSIM of 0.99, which firmly proves the ability of the U-net for denoising.

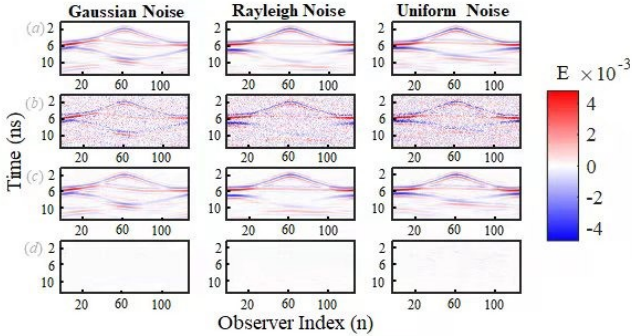

**Fig. 10** Randomly selected examples to demonstrate the denoising performance for three type of noises (1 dB). (a) The ground truth, (b) the electrical field with noise, (c) the output of the U-net and (d) the error between (a) and (c).

In addition, when investigating the performance of complex deep neural networks, ablation research is extensively applied to better understand the behavior of a certain sub-module. After removing a certain part of the system, the necessity of this part can be studied independently by observing the changes in the result. In this research, the predictions w/o the denoising U-net are compared to each other in Fig. 11. It can be concluded that the implement of the denoising U-net significantly improves the inversion accuracy, hence has the prospect of application in practical fields.

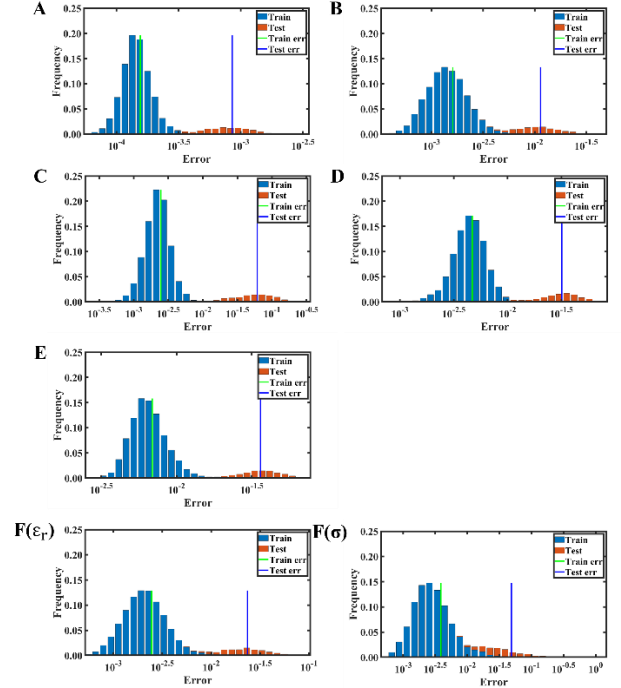

**Fig. 11** The PMSE of the six datasets without the denoising U-net. A: the stratum layer models, B: circles and rings, C: rectangles, D: handwriting numbers and letters from EMNIST, E: multiple scatterers from dataset D, F: lossy scatterers.

### C. Computational acceleration

One of the major motivations behind utilizing TICaN to reconstruct the permittivity and conductivity is that it can yield faster predictions than conventional full-wave inversion algorithms. To this end, the proposed TICaN fully excavates the parallel computing capability of GPUs and thus simultaneously gives predictions with large amount of input data. To quantitatively assess the computational acceleration, we record the calculation time of 3000 samples using the proposed TICaN and the traditional method SOM [38][39] accelerated by FFT [40]. As exhibited in Table 2, it takes 5.3 s on average to predict a specimen for FFT-SOM, which is more than 700 times slower than TICaN. Consequently, it can be concluded that the TICaN has great superiority in computing efficiency over traditional inversion approaches based on continuous iteration, hence can be applied in real-time scenarios.

| **Model** | **Running time (s)** |
|---|---|
| U-net for denoising | 2.1 |
| Efficientunet for inversion | 19.5 |
| TICaN | 21.6 |

## D. Generalization ability

It is of great significance to verify that the developed TICaN has learnt the inherent physical connection of the time-domain scattering fields and the scatterers rather than simply overfitting between the training datasets. Here, the criterion is usually termed as the generalization ability. To measure this ability, one typical approach is to introduce several specimens which has different geometries with the training/testing datasets, and feed them into the proposed architecture. In our studies, the famous "Austria" scatterer is investigated, which is widely encountered in inverse scattering experiments. Fig. 12 summarizes the results of the reconstruction performance, where the ground truth, the retrieved permittivity and the error are shown respectively.

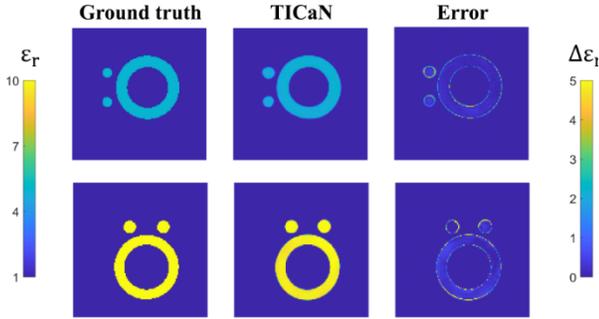

**Fig. 12** The results of the reconstruction performance of the "Austria" scatterer by TICaN.

It can be concluded that the reconstruction results coincident well with the ground truth and the boundary of the scatterer is sharp and clear. It is well known that boundary ambiguity is a common weakness for traditional iterative inversion algorithms, while the proposed TICaN can overcome this defect and achieve a high fidelity. In fact, the average PMSE for the two examples are 0.1163 and 0.1008 respectively, fully indicating the promise generalization ability of the proposed framework.

## IV. CONCLUSION

In this article, a time-domain inversion cascade network (TICaN) is proposed to tackle inverse scattering problems under high-intensity noise, which is the first time for the deep learning technique to be applied to reconstruct the electromagnetic constitute parameters from raw time domain data. It is worth noting that the scatterers investigated in the research contain multitudinous sophisticated datasets, such as the stratum layer, hyperfine structure and lossy medium. After being fully trained, the performance of the TICaN is assessed in terms of accuracy, noise-immunity, computational acceleration, and generalizability. The results have firmly demonstrated that the proposed framework can achieve high-precision inversion under various high-noisy environments. It can be observed that even for scatterers with intricate geometry shapes, hyperfine structure and high contrast, the network can realize super-resolution imaging. Besides, compared to the traditional inversion algorithms, the time consumption of TICaN is reduced significantly, thus the framework offers the potential for converting inverse scattering problems from offline computation to online. More importantly, the proposed TICaN has certain generalization ability in retrieving the unknown constitute parameters (such as the famous "Austria" scatterer) rather than only overfitting between the trained data samples. Considering that the time-domain data is more accessible, we anticipate the proposed TICaN can be extensively adopted in practical real-time scenarios such as biomedical imaging, geological exploration and so on.

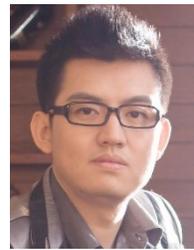

**Qiang Ren** received the B.S. and M.S. degrees both in Electrical Engineering from Beihang University, Beijing, China, and Institute of Acoustics, Chinese Academy of Sciences, Beijing, China in 2008 and 2011, respectively, and the PhD degree in Electrical Engineering from Duke University, Durham, NC, in 2015. From 2016 to 2017, he was a postdoctoral researcher with the Computational Electromagnetics and Antennas Research Laboratory (CEARL) of the Pennsylvania State University, University Park, PA. In Sept 2017, he joined the School of Electronics and Information Engineering, Beihang University, Beijing, China, as an "Excellent Hundred" Associate Professor.

Dr. Ren is the recipient of the Young Scientist Award of 2018 International Applied Computational Electromagnetics Society (ACES) Symposium in China. He serves as the Associate Editor of ACES Journal and Microwave and Optical Technology Letters (MOTL), and also serves as a reviewer for more than 30 journals. His current research interests include numerical modeling methods for complex media, multiscale and multiphysics problems, inverse scattering, deep learning and parallel computing.